# A Predicted Small and Round Heliosphere

**Authors:** Merav Opher[*,2], Abraham Loeb[2], James Drake[3], Gabor Toth[4]

**Affiliations:**

[*]Astronomy Department, Boston University, Boston, MA 02215.

[2]Institute of Theory and Computation, Harvard University, Cambridge, MA

[3]University of Maryland, College Park, MD

[4]University of Michigan, Ann Arbor, MI 48109

*Correspondence to: Merav Opher, mopher@bu.edu

The shape of the solar wind bubble within the interstellar medium, the so-called heliosphere, has been explored over six decades[1-7]. As the Sun moves through the surrounding partially-ionized medium, neutral hydrogen atoms penetrate the heliosphere, and through charge-exchange with the supersonic solar wind, create a population of hot pick-up ions (PUIs). The Termination Shock (TS) crossing by Voyager 2 (V2) data[8] demonstrated that the heliosheath (HS) (the region of shocked solar wind) pressure is dominated by suprathermal particles. Here we use a novel magnetohydrodynamic model that treats the freshly ionized PUIs as a separate fluid from the thermal component of the solar wind. Unlike previous models[9-11], the new model reproduces the properties of the PUIs and solar wind ions based on the New Horizon[12] and V2[8] spacecraft observations. The PUIs charge exchange with the cold neutral H atoms of the ISM in the HS and are quickly depleted. The depletion of PUIs cools the heliosphere downstream of the TS, "deflating" it and leading to a narrower HS and a smaller and rounder shape, in agreement with energetic neutral atom observations by the Cassini spacecraft[7]. The new model, with interstellar magnetic field orientation constrained by the IBEX ribbon[13], reproduces the magnetic field data outside the HP at Voyager 1(V1)[14]. We present the predictions for the magnetic field outside the HP at V2.

The shape of the heliosphere has been explored in the last six decades[1-4]. There was a consensus, since the pioneering work of Baranov & Malama[5], that the heliosphere shape is comet-like. More recently, this standard shape has been challenged by the realization that the solar magnetic field plays a crucial role in funneling the heliosheath solar wind flow into two jet-like structures[6,16]. Cassini's observations of energetic neutral atoms further suggest that the heliosphere has no tail[7]. The crossing of the termination shock (TS) by Voyager 2 (V2)[8] revealed that it is a quasi-perpendicular shock that only weakly heats the thermal plasma while transferring most of the solar wind kinetic energy **into suprathermal particles. The suprathermal particles consist of freshly ionized pickup ions (PUIs) or "core" PUIs and a population of particles with higher energy. PUIs are hot protons created when neutral interstellar hydrogen is ionized in the heliosphere and picked up by the high-speed solar wind. Here we use PUIs as the "core" population. There are also higher energy particles that produce a quiet-time tail of energetic particles that are produced elsewhere in the heliosphere. In this work, we neglect their contribution**. V2 was only able to measure the thermal plasma and measured a downstream temperature of $\sim 10^5$ K. Based on energy conservation, the PUIs are expected to have a downstream temperature of $\sim 10^7$ K.

Previous 3D models for the global structure of heliosphere followed the PUIs and the thermal cold solar wind plasma using a *single-ion* fluid approximation[6,17,18], assuming that the PUIs are immediately mixed into the ambient solar wind plasma. Localized simulations explored the energy conversion across the TS[19-21]. Ref. *22* was the first to treat the PUIs as a separate fluid, assuming no thermal coupling between the two fluids and that they share the same bulk velocity. The first model to treat the interaction with the interstellar medium (ISM) including the PUIs as separate component assumed also that the solar wind protons and PUIs are co-moving[9].

Recent models[10,11] added the solar and interstellar magnetic fields. These global descriptions still treated the solar wind and PUIs as co-moving, and suffer from several other limitations, including a simplified neutral hydrogen atom description[10] and a limited resolution of the heliospheric tail[10,11]. Most importantly, in order to be consistent with the weak shock observed by V2, it is crucial to heat up the PUIs to high temperatures upstream the shock as inferred from extrapolations of the New Horizons data[12].

Here we present a novel 3D magnetohydrodynamic (MHD) model that treats the PUIs as a separate fluid. We solve the full set of fluid equations for both components (including separate energy and momentum equations[23]; see Methods). The model is designed to match the density and temperature of PUIs upstream of the TS consistent with recent observations of New Horizons[12].

New Horizons has recently been making the first direct observations of PUIs in the supersonic solar wind[12]. One of the surprising results is that the PUI temperature is increasing with distance as $r^{0.68}$ and the density of PUIs is decreasing as $r^{-0.6}$, less rapidly than the $r^{-1}$ scaling expected from first order approximations to the PUI mass loading[24]. Based on observations at 30AU and 38AU, the extrapolated temperature and density of PUIs at 90AU are $8.7 \times 10^6$K and $2.2 \times 10^{-4}$cm$^{-3}$. These measurements indicate that the PUI thermal pressure is a substantial fraction of the ram pressure of the solar wind upstream of the TS.

The increase of PUI pressure in the supersonic solar wind could be due to several reasons. Ref. *12* speculates that it could be due to co-rotating interaction regions that merge and drive compression and heating. Due to the uncertain interpretation, we adopt an *ad-hoc* heating of the PUI in the supersonic solar wind to bring their temperature close to $10^7$K upstream of the TS. This value is in agreement with previous work that reproduced the TS crossing[21]. Our model naturally reproduces the value of the density of the PUIs upstream the TS (see Supplementary Table 1, Supplementary Fig. 1).

**We run two models. Model A has the interstellar magnetic field ($B_{ISM}$) in the hydrogen deflection plane (-34°.7 and 57°.9 in ecliptic latitude and longitude, respectively) as in our older single-ion model[6]. Model B uses the $B_{ISM}$ based on the circularity of the IBEX ribbon and the ribbon location[13] (-34°.62 and 47°.3 in ecliptic latitude and longitude, respectively). The strength of the $B_{ISM}$ is 4.4µG for Model A and 3.2µG for Model B.** We map the entire heliosphere with high grid resolution, including the heliospheric tail (with resolution of 3AU throughout the tail for Model A and 2AU for Model B until 400AU and 4AU until 600AU) and 1.5AU across the TS for Model A (and 1.0AU for Model B). We run both models in a time-dependent fashion for 81.3 year's (corresponding to N=400,000 iterations) for Model A and 78 years for Model B (corresponding to N=550,000 iterations). This time is sufficient since it takes a year for the solar wind to reach the TS (100AU) with a velocity of 400km/s. In the heliosheath the speeds are ~ 50km/s so 80 years allows flows to traverse scales twice the size of the heliosphere ~ 400AU. We run the model with a point implicit scheme. Details of the simulations and the two cases are in Methods.

V2 crossed the TS three times (due to radial motions of the TS) and the shock compression ratio was weak (~ 2.3-2.4)[8]. The observed solar wind speed and density were $v_{SW}$ ~ 300km/s and $n_{SW} = 10^{-3}$ cm$^{-3}$. The observed temperature of the solar wind upstream the TS was $T_{SW}$ ~ $10^4$K. This temperature was found by V2 to be roughly constant at distances r > 10AU[25]. This value is larger than expected from the adiabatic expansion of the solar wind. The reason for its high temperature is most likely due to turbulence driven by waves generated by the pick-up process and isotropization of the interstellar PUIs in the solar wind[25,26]. Our model does not include turbulence so the solar wind temperature declines adiabatically with radius (modified by charge exchange- see Supplementary Fig. 1). However, with our choice of the inner boundary value of $T_{SW} = 2.10^4$ K, it reaches approximately the value of $T_{SW} \sim 10^4$ K as observed upstream of the TS by V2.

Our model reproduces the jumps in density, velocity and thermal solar wind temperature as measured by V2 across the TS (Fig. 4). As was shown in local one-dimensional simulations[20], the PUIs carry most of the energy downstream of the TS and the heliosheath thermal pressure is dominated by PUIs and not by the thermal component (Supplementary Fig. 1). Our global model treats the crossing of the TS self-consistently by solving conservative equations for the separate ion fluids (see Methods) that conserve mass, momentum and the hydrodynamic energy. This is a good approximation if the magnetic energy is small relative to the total energy density, which is true for the outer heliosphere. By the fluid nature of our description we are not able to capture kinetic effects such as shock acceleration of PUIs to higher energies which can produce a non-negligible addition to pressure in the HS[30].

The presence of the PUIs as a separate fluid changes the energetics of the global heliosheath (HS) and the overall structure of the heliosphere in two important ways. First, the PUIs weaken the TS by reducing the overall compression across the shock (Fig. 4). This means that the overall power going into the HS from the TS is smaller than in previous models. However, much of the energy that goes into the PUIs is eventually lost due to charge exchange with the interstellar neutrals downstream of the shock. As the PUIs charge exchange, they become energetic neutral atoms (ENAs) and leave the system because the mean-free-path of these particles is greater than the characteristic scale of the HS. This cools the heliosheath much more quickly with distance downstream of the TS compared with the old model that treated the PUI and thermal components as a single fluid, where the loss of PUIs due to charge exchange was not included (Figs. 2B and 2E). **The depletion of PUIs dramatically cools the outer heliosphere, "deflating" it and leading to a smaller and rounder shape than previously predicted.**

As a result of the weaker shock, the magnetic field in the HS just downstream of the TS is weaker than the old single-ion models (Fig. 3D – dash-dotted lines). However, the drop in the PUI pressure leads to compression of the magnetic field further downstream of the TS. In the end, the magnetic field becomes large, not at the shock but further downstream near the HP (Fig. 2F). The strong magnetic field near the HP means that the solar magnetic field continues to play a key role in controlling the overall shape of the heliosphere[6].

Second, the strong gradients of the PUI thermal pressure within the HS (Figure 2B) drive faster flows to the north and south (Fig. 2A and Fig. 2D). As discussed by Drake et al.[16], the HS thickness is controlled by the continuity requirement: plasma flows across the TS must be balanced by flow down the tail within the heliosheath. Stronger flows in the HS therefore reduce the thickness of the HS by deflating the heliospheric bubble and allowing the HS to be compressed by the interstellar medium (Fig. 3D).

The consequence of these two effects is a more "squishable" heliosphere that has a smaller and rounder shape. This global structure is drastically different from the standard picture of a long heliosphere with a comet-like tail that extends to thousands of AUs (Fig. 1B). The distance from the sun to the heliopause in the new round heliosphere is nearly the same in all directions. This new rounder and smaller shape is in agreement with the ENA observations by Cassini spacecraft[7]. **The Cassini ENAs are produced by more energetic particles than the ones in our model, the "core PUIs". One can observe farther down the tail at the higher** **energies measured by Cassini. If the high energy particles measured by Cassini imply a round heliosphere, the PUIs must also occupy a volume that is round since their loss mean-free-path is shorter than that of the energetic ions.**

Our model predicts that PUIs stream with a higher velocity along magnetic field lines away from the nose of the HS, than the solar wind ions (Fig. 3E). This large velocity is driven by the large drop in the PUI pressure towards the flanks (Fig. 3F). The motion of the thermal and PUI fluids perpendicular to the local magnetic field is controlled by the local electric field and is therefore the same for both species except in regions with large perpendicular gradients in pressure such as at the TS[21]. The perpendicular velocity of both species is therefore nearly the same over most of the region downstream of the TS (see Supplementary Fig. 4). Along the magnetic field, however, the ion fluids are decoupled and can attain significantly different velocities. In reality, two-stream instabilities restrict the relative ion velocities parallel to the magnetic field. These instabilities are a kinetic phenomenon that cannot be represented in multi-ion MHD. We therefore use a nonlinear artificial friction source term in the momentum equation to limit the relative velocities to realistic values (as in Ref. *23*). This *ad-hoc* artificial friction term in practice limits the velocity difference to the local Alfven speed (see Methods). Typically, the velocity difference between PUIs and SW is around 40km/s and is field aligned (Fig. 3E).

The heliosheath is dominated by thermal pressure almost all the way to the heliopause (Fig. 2F and Supplementary Fig. 2). Only near the heliopause is the magnetic pressure is larger than the total thermal pressure. The temperature of the heliosheath is dominated by the PUIs and is around 2 keV (Figs. 3G and 3H). The density of PUIs quickly decreases with distance downstream of the TS in the heliosheath (Fig. 3A). The spatial profile of the PUIs will affect interpretation of the ENA maps measured at IBEX in the energy range of 1-4keV.

**Both Voyager spacecraft crossed the HP at roughly the same distance (~122 AU for V1 and 119 AU for V2[15]). Model B with $B_{ISM}$ along the center of the IBEX ribbon[13], reproduces the V1 observations of the magnetic field outside the heliosphere (Figure 5) while Model A does not. The distances to the HP in the V1 and V2 directions in Model B are comparable although both exceed the distances measured by the two spacecraft (Table 1).**

**All present models of the global heliosphere yield thicknesses of the HS that are substantially larger than measured by the Voyagers. The thickness of the HS in the new multi-ion MHD model is significantly reduced as compared to the single ion case (Table 1) and other models. This because the TS moves outward and the HP moves inward when compared to the simplified case where the PUIs and thermal plasma are treated as a single fluid (Fig. 2A; Fig. 2D; and Table 1). Again, this is a result of the effective "deflation" of the heliospheric bubble due to the charge exchange losses of the PUIs. This effect was also seen in previous models that treated the PUI as a separate component[9].**

In addition, in these new simulations, the thermal pressure of PUIs upstream the TS is greater than in earlier single ion models (Supplementary Fig. 2) (around 30% of the ram pressure, $2.6x10^{-14}$ ergs/cm$^3$) while the solar wind thermal pressure is an order of magnitude smaller ( $1.1x\ 10^{-16}$ ergs/cm$^3$) than the ram pressure –( $8.3x10^{-14}$ ergs/cm$^3$). The increased pressure of the PUIs that pushes the TS outward in the new model. **Model A predicts that at V1 the distance to the TS is 105AU±3AU while at V2 it is 100AU±3AU, so the asymmetry in the TS in our new multi-ion model is reduced from previous single ion models[28]. The distances to the TS for Model B are 102AU± 3AU and 98AU± 3AU at V1 and V2 respectively, and are still somewhat larger than the observations (95AU at V1 and 85AU at**

V2). However, the location of the TS can be adjusted by slightly altering the ISM conditions (increased pressure of the ISM reduces the TS radius). In contrast, it is much more difficult to control the thickness of the HS in a model. An important scientific result of the new multi-ion model on top of the change in shape is the significant reduction in the thickness of the HS.

Our new model does not include the solar cycle variation of the solar wind. However, time-dependent simulations[29] show that the TS only fluctuates by ±10AU with the solar cycle, while fluctuations of the HP distance are only ~3-4 AU. Thus, solar cycle variability can not explain the continuing discrepancy between the thin HS measured by the Voyagers and the global models.

An important extension of this work would be to include not only the PUIs created in the supersonic solar wind (which peak around 1-3keV) but the higher energy particles such as anomalous cosmic rays (ACRs) that are measured by Voyager 1 from 30keV up to several MeV's. While none of the global models include ACRs, the diffusive loss of cosmic rays through the heliopause was predicted to shift the positions of the TS and HP by around ~ 5AU[30].

Finally, it can be seen that, although the heliosphere has a short tail, the ISM down the tail is affected far downstream (Fig. 2C). The disturbance is due to the mixing between the HS and ISM plasma as the lobes become turbulent and magnetic fields reconnect. For example, there is substantial material between the lobes sitting on field lines open to the ISM[31]. Potentially that region could contain HS PUIs that could undergo local acceleration. The effect of this turbulent domain on ENA maps remains to be explored. This region could influence the models that are used to probe the heliospheric tail such as Lyman alpha emission[32] and TeV cosmic rays[33].

Future remote sensing and *in-situ* measurements will be able to test the reality of a rounder heliosphere. **In Figure 5 we show our prediction for the interstellar magnetic field ahead of the heliosphere at V2.** In addition, future missions such as the Interstellar Mapping and Acceleration Probe will return ENA maps at higher energies than present missions and so will be able to explore ENAs coming from deep into the heliospheric tail. Thus, further exploration of the global structure of the heliosphere will be forthcoming and will put our model to the test.

**Methods**
**Description of the Governing Equations.** Our model has two ions, solar wind and PUIs interacting through charge exchange with neutral H atoms. The neutral H atoms are described in a multi-fluid treatment. There are four neutral populations, each reflecting the properties of the plasma between the different heliospheric boundaries[28].

The model assumes "cold electron" approximation, i.e., that there are no suprathermal electrons. This is in agreement with the observations[8]. With $n_{SW}$ and $n_{PUI}$ being, respectively the number density of the thermal solar wind protons and the PUIs, from charge neutrality we have,

$$n_e = n_{SW} + n_{PUI} \quad (1)$$

Assuming that $T_e=T_{SW}$, where $T_{SW}$ is the proton temperature the solar wind thermal pressure is

$$p_{SW} = (n_{SW}T_{SW} + n_e T_e)k_B = (2n_{SW} + n_{PUI})T_{SW}k_B$$

The PUI pressure is $p_{PUI} = n_{PUI}T_{PUI}k_B$.

We solve the multi-fluid set of equations (as in *23*; *35*) for the solar wind and PUIs modified to include source terms due to charge exchange as in ref. *36*,

$$\frac{\partial \rho_{SW}}{\partial t} + \vec{\nabla} \cdot (\rho_{SW}\vec{u}_{SW}) = S_{\rho_{SW}} \quad (2)$$

$$\frac{\partial \rho_{PUI}}{\partial t} + \vec{\nabla} \cdot (\rho_{PUI}\vec{u}_{PUI}) = S_{\rho_{PUI}} \quad (3)$$

$$\frac{\partial (\rho_{SW}\vec{u}_{SW})}{\partial t} + \vec{\nabla} \cdot \left(\rho_{SW}\vec{u}_{SW}\vec{u}_{SW} + p_{SW}\overleftrightarrow{I}\right) - \frac{\rho_{SW}}{m_p}(\vec{u}_{SW} - \vec{u}_+)\times\vec{B} - \frac{\rho_{SW}}{n_e e}\vec{J}\times\vec{B} = S_{M_{SW}} \quad (4)$$

$$\frac{\partial (\rho_{PUI}\vec{u}_{PUI})}{\partial t} + \vec{\nabla} \cdot \left(\rho_{PUI}\vec{u}_{PUI}\vec{u}_{PUI} + p_{PUI}\overleftrightarrow{I}\right) - \frac{\rho_{PUI}}{m_p}(\vec{u}_{PUI} - \vec{u}_+)\times\vec{B} - \frac{\rho_{PUI}}{n_e e}\vec{J}\times\vec{B} = S_{M_{PUI}} \quad (5)$$

$$\frac{\partial \varepsilon_{SW}}{\partial t} + \vec{\nabla} \cdot [(\varepsilon_{SW} + p_{SW})\vec{u}_{SW})] - \frac{\rho_{SW}}{m_p}\vec{u}_{SW} \cdot (\vec{u}_{SW} - \vec{u}_+)\times\vec{B} - \frac{\rho_{SW}}{n_e e}\vec{u}_{SW}\cdot\vec{J}\times\vec{B} = S_{\varepsilon_{SW}}$$

(6)

$$\frac{\partial \mathcal{E}_{PUI}}{\partial t} + \vec{\nabla} \cdot [(\mathcal{E}_{PUI} + p_{PUI})\vec{u}_{PUI})] - \frac{\rho_{PUI}}{m_p}\vec{u}_{PUI} \cdot (\vec{u}_{PUI} - \vec{u}_+) \times \vec{B} - \frac{\rho_{PUI}}{n_e e}\vec{u}_{PUI} \cdot \vec{J} \times \vec{B} = S_{\mathcal{E}_{PUI}} + H,$$

(7)

where $\vec{u}_+ = \frac{\rho_{SW}\vec{u}_{SW} + \rho_{PUI}\vec{u}_{PUI}}{\rho_{SW} + \rho_{PUI}}$ is the charge-averaged ion velocity and the source terms, S represent the mass, momentum, and energy sources respectively due to charge exchange[36]. In equation 7, we include in the possibility that the PUIs are heated in the supersonic solar wind with the variable "H". This is because observations by New Horizons[12] show that the PUIs are heated as a function of distance.

The radiation pressure and the gravity are assumed to perfectly cancel each other out. Ionization processes such as photoionization and electron-impact ionization are also neglected. These processes play a much lesser role than charge exchange at larger radii (R > 30 AU).

The neutrals H atoms are described as 4 different populations having the characteristics of different regions of the heliosphere[28]. The four populations of neutral hydrogen atoms have different origins: atoms of interstellar origin represent population 4; Population 1 is created by charge exchange in the region behind the interstellar bow shock (or slow bow shock[37]) and Populations 3 and 2 originate from the supersonic solar wind and the heliosheath, respectively. All four populations "i" index, are described by separate systems of the Euler equations with the corresponding source terms describing the ion-neutral (both the solar wind and PUIs) charge exchange process,

$$\frac{\partial \rho_H(i)}{\partial t} + \vec{\nabla} \cdot (\rho_H \vec{u}_H) = S_{\rho_H}(i) \qquad (8)$$

$$\frac{\partial \rho_H \vec{u}_H}{\partial t} + \vec{\nabla} \cdot (\rho_H \vec{u}_H \vec{u}_H + p_{PUIH}\overleftrightarrow{I}) = S_{M_H}(i) \qquad (9)$$

$$\frac{\partial \mathcal{E}_H}{\partial t} + \vec{\nabla} \cdot [(\mathcal{E}_H + p_H)\vec{u}_H)] = S_{\mathcal{E}_H}(i) \qquad (10)$$

Source Terms

We describe next considering our multi-fluid description of the neutrals, which charge exchange processes occur. In the supersonic solar wind (what we refer as Region 3), the following charge-exchange processes occur,

$$p_0 + H_1 \rightarrow p_1 + H_3$$
$$p_0 + H_2 \rightarrow p_1 + H_3$$
$$p_0 + H_3 \rightarrow p_0 + H_3$$
$$p_0 + H_4 \rightarrow p_1 + H_3$$

and

$$p_1 + H_1 \rightarrow p_1 + H_3$$
$$p_1 + H_2 \rightarrow p_1 + H_3$$
$$p_1 + H_3 \rightarrow p_0 + H_3$$
$$p_1 + H_4 \rightarrow p_1 + H_3$$

where $p_0$ is the solar wind proton, $p_1$ the PUI and $H_1$, $H_2$, $H_3$, $H_4$ are, respectively, the neutrals of population 1, 2, 3 and 4.

Outside of region 3, the following charge exchange processes occur,

$$p_0 + H_1 \rightarrow p_0 + H_2$$

$$p_0 + H_2 \rightarrow p_0 + H_2$$
$$p_0 + H_3 \rightarrow p_0 + H_2$$
$$p_0 + H_4 \rightarrow p_0 + H_2$$

and

$$p_1 + H_1 \rightarrow p_0 + H_2$$
$$p_1 + H_2 \rightarrow p_0 + H_2$$
$$p_1 + H_3 \rightarrow p_0 + H_2$$
$$p_1 + H_4 \rightarrow p_0 + H_2$$

Density Source Terms

In region 3, in the supersonic solar wind, the density source term for the solar wind protons is

$$S_{\rho SW} = -\sum_{i=1}^{4} \rho_{SW} n_H(i) U^*(i) \sigma_{NSW}(i) + \rho_{SW} n_H(3) U^*(3) \sigma_{NSW}(3) + \rho_{PUI} n_H(3) U^*(3) \sigma_{NPUI}(3) \tag{11}$$

and for the PUIs is

$$S_{\rho PUI} = \sum_{i=1}^{4} \rho_{SW} n_H(i) U^*(i) \sigma_{NSW}(i) - \rho_{SW} n_H(3) U^*(3) \sigma_{NSW}(3) - \rho_{PUI} n_H(3) U^*(3) \sigma_{NPUI}(3) \tag{12}$$

The source terms in density of the neutral populations $i$=1,2,4 and Pop 3 are:

$$S_{\rho H}(i) = -\rho_{SW} n_H(i) U^*(i) \sigma_{NSW}(i) - \rho_{PUI} n_H(i) U^*(i) \sigma_{NPUI}(i) \tag{13}$$
$$S_{\rho H}(3) = \sum_{i=1}^{4} \rho_{SW} n_H(i) U^*(i) \sigma_{NSW}(i) + \sum_{i=1}^{4} \rho_{PUI} n_H(i) U^*(i) \sigma_{NPUI}(i) - \rho_{SW} n_H(3) U^*(3) \sigma_{NSW}(3) - \rho_{PUI} n_H(3) U^*(3) \sigma_{NPUI}(3) \tag{14}$$

In region 2, in the heliosheath, the density source term for the solar wind protons is

$$S_{\rho SW} = \sum_{i=1}^{4} \rho_{PUI} n_H(i) U^*(i) \sigma_{NPUI}(i) \tag{15}$$

and for the PUIs is

$$S_{\rho PUI} = -\sum_{i=1}^{4} \rho_{PUI} n_H(i) U^*(i) \sigma_{NPUI}(i) \tag{16}$$

The density source terms of the neutral populations $i$=1, 3, 4 and Pop 2 are:

$$S_{\rho H}(i) = -\rho_{SW} n_H(i) U^*(i) \sigma_{NSW}(i) - \rho_{PUI} n_H(i) U^*(i) \sigma_{NPUI}(i) \tag{17}$$
$$S_{\rho H}(2) = \sum_{i=1}^{4} \rho_{SW} n_H(i) U^*(i) \sigma_{NSW}(i) + \sum_{i=1}^{4} \rho_{PUI} n_H(i) U^*(i) \sigma_{NPUI}(i) \tag{18}$$

Momentum Source Terms

In region 3, in the supersonic solar wind, the momentum source term for the solar wind protons is

$$S_{M_{SW}} = -\sum_{i=1}^{4} \rho_{SW} n_H(i) U_M^*(i) \sigma_{SW}(i) U_{SW} + \rho_{SW} n_H(3) U_M^*(3) \sigma_{SW}(3) U_H(3) + \rho_{PUI} n_H(3) U_M^*(3) \sigma_{PUI}(3) U_H(3) \quad (19)$$

and for the PUIs is

$$S_{M_{PUI}} = \sum_{i=1}^{4} \rho_{PUI} n_H(i) U_M^*(i) \sigma_{SW}(i) \Delta U_{PUI-H}(i) + \sum_{i=1}^{4} \rho_{SW} n_H(i) U_M^*(i) \sigma_{SW}(i) U_H(i) - \rho_{SW} n_H(3) U_M^*(3) \sigma_{SW}(3) U_H(3) - \rho_{PUI} n_H(3) U_M^*(3) \sigma_{PUI}(3) U_H(3) \quad (20)$$

The momentum source terms of the neutral populations $i=1, 2, 4$ and Pop 3 are:

$$S_{M_H}(i) = -\rho_{SW} n_H(i) U_M^*(i) \sigma_{SW}(i) U_H(i) - \rho_{PUI} n_H(i) U_M^*(i) \sigma_{SW}(i) U_H(i) \quad (21)$$
$$S_{M_H}(3) = \sum_{i=1}^{4} \rho_{SW} n_H(i) U_M^*(i) \sigma_{SW}(i) U_{SW} + \sum_{i=1}^{4} \rho_{PUI} n_H(i) U_M^*(i) \sigma_{PUI}(i) U_{PUI} - \rho_{SW} n_H(3) U_M^*(3) \sigma U_H(3) - \rho_{PUI} n_H(3) U_M^*(3) \sigma U_H(3) \quad (22)$$

In region 2, in the heliosheath, the momentum source term for the solar wind protons is

$$S_{M_{SW}} = \sum_{i=1}^{4} \rho_{SW} n_H(i) U_M^*(i) \sigma_{SW}(i) \Delta U_{SW-H}(i) + \sum_{i=1}^{4} \rho_{PUI} n_H(i) U_M^*(i) \sigma_{PUI}(i) U_H(i) \quad (23)$$

and for the PUIs is

$$S_{M_{PUI}} = -\sum_{i=1}^{4} \rho_{PUI} n_H(i) U_M^*(i) \sigma_{PUI}(i) U_{PUI} \quad (24)$$

The momentum source terms of the neutral populations $i=1, 3, 4$ and Pop 2 are:

$$S_{M_H}(i) = -\rho_{SW} n_H(i) U_M^*(i) \sigma U_H(i) - \rho_{PUI} n_H(i) U_M^*(i) \sigma U_H(i) \quad (25)$$
$$S_{M_H}(2) = \sum_{i=1}^{4} \rho_{SW} n_H(i) U_M^*(i) \sigma_{SW}(i) U_{SW} + \sum_{i=1}^{4} \rho_{PUI} n_H(i) U_M^*(i) \sigma_{PUI}(i) U_{PUI} - \rho_{SW} n_H(2) U_M^*(2) \sigma_{SW}(2) U_H(2) - \rho_{PUI} n_H(2) U_M^*(2) \sigma_{PUI}(2) U_H(2) \quad (26)$$

Energy Source Terms
In region 3, in the supersonic solar wind, the energy source term for the solar wind protons is

$$S_{\mathcal{E}_{SW}} = -\sum_{i=1}^{4} 0.5 \rho_{SW} n_H(i) U_M^*(i) \sigma_{SW}(i) U_{SW}^2 - \rho_{SW} n_H(i) U_M^*(3) \sigma_{SW}(3) U_{thSW} + 0.5 \rho_{SW} n_H(3) U_M^*(3) \sigma_{SW}(3) U_H^2(3) + \rho_{SW} n_H(3) U_M^*(3) \sigma_{SW}(3) U_{th}(3) + 0.5 \rho_{PUI} n_H(3) U_M^*(3) \sigma_{PUI}(3) U_H^2(3) + \rho_{PUI} n_H(3) U_M^*(3) \sigma_{PUI}(3) U_{th}(3) \quad (27)$$

and for the PUIs is

$$S_{\mathcal{E}_{PUI}} = \sum_{i=1}^{4} 0.5 \rho_{PUI} n_H(i) U_M^*(i) \sigma_{PUI}(i) \left( U_H^2(i) - U_{PUI}^2 \right) + \rho_{PUI} n_H(i) U_M^*(i) \sigma_{PUI}(i) (U_{th}(i) - U_{thPUI}) - 0.5 \rho_{PUI} n_H(3) \sigma_{PUI}(3) U_M^*(3) U_H^2(3) - \rho_{PUI} n_H(3) U_M^*(3) \sigma_{PUI}(3) U_{th}(3) + \sum_{i=1}^{4} 0.5 \rho_{SW} n_H(i) U_M^*(i) \sigma_{SW}(i) U_H^2(i) +$$

$$\rho_{SW} n_H(i) U_M{}^*(i) \sigma_{SW}(i) U_{th}(i) - 0.5 \rho_{SW} n_H(3) U_M{}^*(3) \sigma_{SW}(3) U_H^2(3) - $$
$$\rho_{SW} n_H(3) U_M{}^*(3) \sigma_{SW}(3) U_{th}(3) \tag{28}$$

The energy source terms of the neutral populations i=1, 2, 4 and Pop 3 are:

$$S_{\varepsilon_H}(i) = -0.5\rho_{SW} n_H(i) U_M{}^*(i) \sigma_{SW}(i) U_H^2(i) - \rho_{SW} n_H(i) U_M{}^*(i) \sigma_{SW}(i) U_{th}(i) +$$
$$-0.5\rho_{PUI} n_H(i) U_M{}^*(i) \sigma_{PUI}(i) U_H^2(i) - \rho_{PUI} n_H(i) U_M{}^*(i) \sigma_{PUI}(i) U_{th}(i)$$
(29)

$$S_{\varepsilon_H}(3) = -0.5\rho_{SW} n_H(3) U_M{}^*(3) \sigma_{SW}(3) U_H^2(3) - \rho_{SW} n_H(3) U_M{}^*(3) \sigma_{SW}(i) U_{th}(3) +$$
$$-0.5\rho_{PUI} n_H(3) U_M{}^*(3) \sigma_{PUI}(3) U_H^2(3) - \rho_{PUI} n_H(3) U_M{}^*(3) \sigma_{PUI}(3) U_{th}(3) +$$
$$\sum_{i=1}^{4}[\rho_{SW} n_H(i) U_M{}^*(i) \sigma_{SW}(i) U_{SW}^2 + \rho_{SW} n_H(i) U_M{}^*(i) \sigma_{SW}(i) U_{thSW}] +$$
$$\sum_{i=1}^{4}[\rho_{PUI} n_H(i) U_M{}^*(i) \sigma_{PUI}(i) U_{PUI}^2 + \rho_{PUI} n_H(i) U_M{}^*(i) \sigma_{PUI}(i) U_{thPUI}] \tag{30}$$

In region 2, in the heliosheath, the energy source term for the solar wind protons is

$$S_{\varepsilon_{SW}} = \sum_{i=1}^{4}[0.5\rho_{SW} n_H(i) U_M{}^*(i) \sigma_{SW}(i)(U_H^2(i) - U_{SW}^2) + \rho_{SW} n_H(i) U_M{}^*(i) \sigma_{SW}(i)(U_{th}(i) -$$
$$U_{thSW})] + \sum_{i=1}^{4}[0.5\rho_{PUI} n_H(i) U_M{}^*(i) \sigma_{PUI}(i) U_H^2(i) + \rho_{PUI} n_H(i) U_M{}^*(i) \sigma_{PUI}(i) U_{th}(i)]$$
(31)

and for the PUIs is

$$S_{\varepsilon_{PUI}} = \sum_{i=1}^{4}[-0.5\rho_{PUI} n_H(i) U_M{}^*(i) \sigma_{PUI}(i) U_{PUI}^2 - \rho_{PUI} n_H(i) U_M{}^*(i) \sigma_{PUI}(i) U_{thPUI}] \tag{32}$$

The energy source terms of the neutral populations *i*=1, 3, 4 and Pop 2 are:

$$S_{\varepsilon_H}(i) = -0.5\rho_{SW} n_H(i) U_M{}^*(i) \sigma_{SW}(i) U_H^2(i) - \rho_{SW} n_H(i) U_M{}^*(i) \sigma_{SW}(i) U_{th}(i) -$$
$$0.5\rho_{PUI} n_H(i) U_M{}^*(i) \sigma_{PUI}(i) U_H^2(i) - \rho_{PUI} n_H(i) U_M{}^*(i) \sigma_{PUI}(i) U_{th}(i) \tag{33}$$

$$S_{\varepsilon_H}(2) = -0.5\rho_{SW} n_H(2) U_M{}^*(2) \sigma_{SW}(2) U_H^2(2) - \rho_{SW} n_H(2) U_M{}^*(2) \sigma_{SW}(2) U_{th}(2) -$$
$$0.5\rho_{PUI} n_H(2) U_M{}^*(2) \sigma_{PUI}(2) U_H^2(2) - \rho_{PUI} n_H(2) U_M{}^*(2) \sigma_{PUI}(2) U_{th}(2) +$$
$$\sum_{i=1}^{4}[\rho_{SW} n_H(i) U_M{}^*(i) \sigma_{SW}(i) U_{SW}^2 + \rho_{SW} n_H(i) U_M{}^*(i) \sigma_{SW}(i) U_{th}(i)] +$$
$$\sum_{i=1}^{4}[\rho_{PUI} n_H(i) U_M{}^*(i) \sigma_{PUI}(i) U_{PUI}^2 + \rho_{PUI} n_H(i) U_M{}^*(i) \sigma_{PUI}(i) U_{thPUI}] \tag{34}$$

In the source terms the following terms appear, where the index "*i*" refers to each population of neutrals 1, 2, 3, or 4. $U_{thSW}$ is the thermal speeds of the solar wind and $U_{thPUI}$ the thermal speed of PUIs:

$$U^*{}_{SW}(i) = \sqrt{\frac{4}{\pi}(w_{SW}^2 + w_H^2(i)) + (\Delta U_{SW-H}(i))^2}, \quad U_{thSW}^2 = \frac{2k_B T_{SW}}{m_p}, \quad U_{th}^2(i) = \frac{2k_B T_H(i)}{m_p},$$

$$U_{thPUI}^2 = \frac{2k_B T_{PUI}}{m_p}, \quad U^*{}_{PUI}(i) = \sqrt{\frac{4}{\pi}(w_{PUI}^2 + w_H^2(i)) + (\Delta U_{PUI-H}(i))^2},$$

$$\Delta U_{SW-H}(i) = \sqrt{\vec{u}_H(i) - \vec{u}_{SW}}; \quad \Delta U_{PUI-H}(i) = \sqrt{\vec{u}_H(i) - \vec{u}_{PUI}},$$

$$U^*_{M-PUI}(i) = \sqrt{\frac{64}{9\pi}(w_{PUI}^2 + w_H^2(i)) + (\Delta U_{PUI-H}(i))^2},$$

$$U^*_{M-SW}(i) = \sqrt{\frac{64}{9\pi}(w_{SW}^2 + w_H^2(i)) + (\Delta U_{SW-H}(i))^2}.$$

The cross sections are from ref. *39*

$$\sigma_{SW}(i) = (2.2835 \times 10^{-7} - 1.062 \times 10^{-8} \ln(U^*_{M-SW}(i) * 100))^2 \times 10^{-4} \, cm^2$$
$$\sigma_{NSW}(i) = (2.2835 \times 10^{-7} - 1.062 \times 10^{-8} \ln(U^*_{SW}(i) * 100))^2 \times 10^{-4} \, cm^2$$

$$\sigma_{PUI}(i) = (2.2835 \times 10^{-7} - 1.062 \times 10^{-8} \ln(U^*_{M-PUI}(i) * 100))^2 \times 10^{-4} \, cm^2$$
$$\sigma_{NPUI}(i) = (2.2835 \times 10^{-7} - 1.062 \times 10^{-8} \ln(U^*_{PUI}(i) * 100))^2 \times 10^{-4} \, cm^2$$

**PUIs Heating Source Term.** The *ad-hoc* heating source term H was chosen to as

$$H = \rho_{PUI}(T_{PUI}(K) - 10^7)(r(AU) - 30.) * 10. \tag{35}$$

only in the supersonic solar wind, where r is the radius and $\rho_{PUI}$ and $T_{PUI}$ are, respectively, the density and temperature of PUIs. This ad-hoc heating brought the temperature of the PUI to $10^7$K upstream the termination shock.

**Numerical Models.**

The inner boundary of our domain is a sphere at 30AU and the outer boundary is at x = ±1500AU, y = ± 1500AU, z = ±1500AU for Model A and for Model B, x = ±1500AU, y = ± 2000AU, z = ±2000AU. We increased the grid size for Model B to capture the slow bow shock[37] that forms along the plane that contains the interstellar magnetic field and interstellar velocity. Parameters of the solar wind at the inner boundary at 30AU were: $v_{SW}$ = 417 km/s, $n_{SW}$ = 8.74 x $10^{-3}$ cm$^{-3}$, $T_{SW}$ = 1.087 x $10^5$ K (OMNI solar data; http://omniweb.gsfc.nasa.gov/). The magnetic field is given by the Parker spiral magnetic field with $B_{SW}$ = 7.17x$10^{-3}$ nT at the equator. We use a monopole configuration for the solar magnetic field. This description while capturing the topology of the field line does not capture its change of polarity with solar cycle or across the heliospheric current sheet. This choice, however, minimizes artificial reconnection effects, especially in the heliospheric current sheet. In our simulation, we assume that the magnetic axis is aligned with the solar rotation axis. The solar wind flow at the inner boundary is assumed to be spherically symmetric. For the interstellar plasma, we assume: $v_{ISM}$ = 26.4 km/s, $n_{ISM}$ = 0.06 cm$^{-3}$, $T_{ISM}$= 6519 K. The number density of H atoms in the interstellar medium is $n_H$ = 0.18 cm$^{-3}$, the velocity and temperature are the same as for the interstellar plasma. The coordinate system is such that Z-axis is parallel to the solar rotation axis, X-axis is 5° above the direction of interstellar flow with Y completing the right-handed coordinate system. The strength of the $B_{ISM}$ in the model is 4.4µG for Model A and 3.2µG for Model B. The orientation of $B_{ISM}$ continues to be debated in the literature. For Model A we use $B_{ISM}$ in the hydrogen deflection plane (-34°.7 and 57°.9 in ecliptic latitude and longitude, respectively) consistent with the measurements of deflection of He atoms with respect to the H atoms[38] and for Model B we use $B_{ISM}$ used in works that constrain the orientation of $B_{ISM}$ based on the circularity of the IBEX ribbon and the ribbon location[13] (-34°.62 and 47°.3 in ecliptic latitude and longitude, respectively).

We assume in the inner boundary (30AU) values of density of the PUIs ($n_{pui} = 9.45\times10^{-4}$ cm$^{-3}$) such that the value upstream of the TS at V2 correspond to the predicted value by New Horizon (Fig. 4) (see Supplementary Table 1). The value of the temperature of the PUI chosen at the inner boundary was $T_{PUI} = 8.2\times10^6$ K. We introduce an ad-hoc heating of the PUI, only in the supersonic solar wind, to bring their temperature to $10^7$K upstream the TS as predicted by New Horizon. The value of the speed of the PUIs in the inner boundary is the same as the solar wind $v_{PUI} = 417$ km/s.

Along the magnetic field, the PUI and solar wind fluids are decoupled and can attain significantly different ion velocities in a cold electron approximation. In reality, two-stream instabilities physically restrict the relative ion velocities parallel to the magnetic field. This two-stream instability is a kinetic phenomenon that cannot be represented in multi-ion MHD, therefore (23) used a nonlinear artificial friction source term in the momentum equation to limit the relative velocities to realistic values,

$$S_M^{friction} = \frac{\rho_{PUI}}{\tau_c}(\vec{u}_{PUI} - \vec{u}_{SW})\left(\frac{|\vec{u}_{PUI}-\vec{u}_{SW}|}{u_c}\right)^{\alpha_c} \qquad (36)$$

where $\tau_c$ is the relaxation time scale, $u_c$ is the cutoff velocity, and $\alpha_c$ is the cutoff exponent. Here we used $\tau_c = 10^6 s$, $\alpha_c = 4$ and $u_c$ is set to the local Alfvén speed using the total ion mass density.

**Regarding the Perpendicular Speeds of PUIs and Solar Wind.** The dominant terms in the PUI and SW momentum equations throughout most of the heliosphere are the terms proportional to $(\vec{u}_+)\times\vec{B}$ and $(\vec{u}_{SW})\times\vec{B}$ or $(\vec{u}_{PUI})\times\vec{B}$ where the first is basically the perpendicular electric field. These terms therefore typically balance so the SW and PUI velocities are equal and given by the $\vec{E}\times\vec{B}$ drift. However, in regions where the local gradients in magnetic field or pressure are large such as at the termination shock other terms in the momentum equation can be significant and as a consequence the velocity of the ions can differ from the local $\vec{E}\times\vec{B}$ drift and therefore differ from each other. For example, large gradients of PUI pressure can make the perpendicular speeds of the PUIs different than the solar wind ions. The term responsible for that, in the momentum equation Eq. (5) is $\nabla p_{PUI}$. Comparing that term with $\vec{u}\times\vec{B}$ the ratio is

$$\frac{\nabla p_{PUI}}{ne\vec{u}\times\vec{B}} \sim \frac{v_{diamag}}{U_{flow}} \sim \frac{v_{th(PUI)}}{U_{flow}}\frac{r_L}{L_p} \qquad (37)$$

where $r_L = \frac{mv_{th(PUI)}}{|q|B}$ is the Larmor radius for the PUI; $L_p$ the length of the gradient of pressure. $v_{diamag}$ and $v_{th(PUI)}$ are, respectively the diamagnetic and thermal speeds of the PUIs,

$$v_{diamg} = \frac{v_{th(PUI)}}{L_p}\frac{mv_{th(PUI)}}{|q|B} \sim \frac{p_{PUI}}{n_{PUI}}\frac{1}{|q|BL_p}$$

The ratio in Eq. (37), $\frac{v_{th(PUI)}}{U_{flow}} \sim 7$ from mid heliosheath to ~ 30 near the heliopause. $V_{th} \sim 4\times10^2$ km/s and the $U_{flow} \sim 60$ km/s. The Larmor radii $r_L \sim 1.02\times10^{-3}$ (T(K))$^{1/2}$B$^{-1}$ km with B ~ 0.34nT in the Heliosheath and T ~ $10^7$ K of PUI

$$r_L \sim 2x10^{-3} \ AU$$

The PUI pressure drops length in the heliosheath is $L_p \sim 25 AU$. One can see that $\frac{\nabla p_{PUI}}{ne\vec{u}\times\vec{B}} \sim 6x10^{-4}$ and the perpendicular speeds for the PUIs and solar wind ions should be the same. Fig S4 shows that the perpendicular speeds are, indeed, the same everywhere in the heliosheath. At the Termination Shock, as shows in ref. *21*, $L_p$ is small ($<r_L$) there should be a difference in the perpendicular speeds in PUI and solar wind.

**Acknowledgments:** The authors would like to thank the staff at NASA Ames Research Center for the use of the Pleiades supercomputer under the award SMD-16-7616 and SMD-18-1875 and especially the staff Ms. Nancy Carney. M. O. acknowledge discussions with Mr. Adam Michael, Mr. Marc Korenbleuth. M. O. and J. D. was partially supported by NASA Grant NNH13ZDA001N-GCR and NNX14AF42G. A. L. acknowledges support from the Breakthrough Prize Foundation.

**Author contributions:** M. O. performed the numerical simulations with guidance and collaboration from G. T. The scientific analysis and discussion of the results were done by all the authors. The manuscript was reviewed and edited by all the authors. **Competing interests:** There are no competing interests to declare. **Data and materials availability:** Our model is the OH module of SWMF is available at http://csem.engin.umich.edu/tools/swmf/.

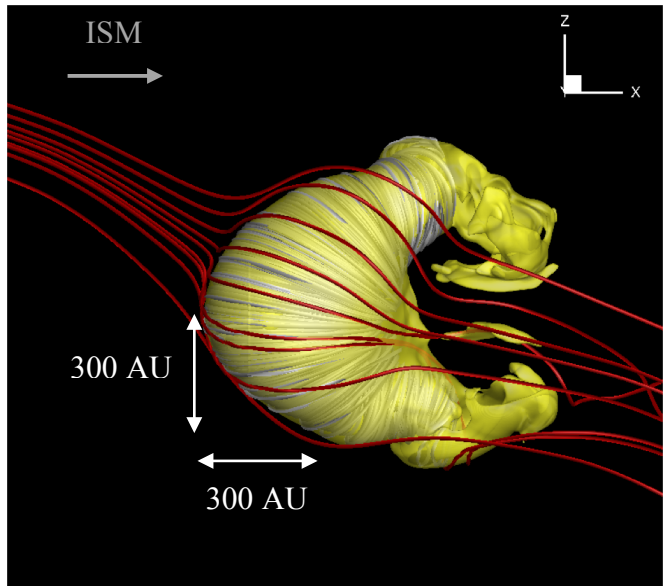 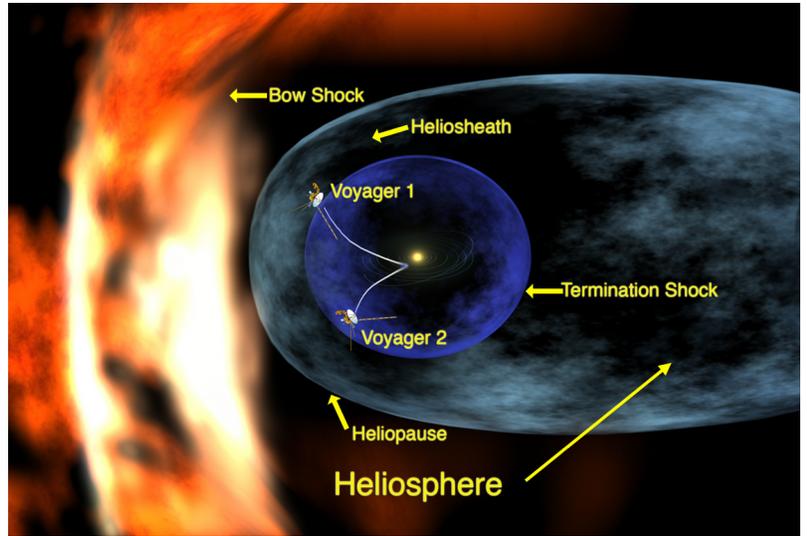

(A)                                                                                      (B)

**Figure 1. The New Heliosphere.** (A) The Heliopause is shown by the yellow surface (Case A) defined by solar wind density = 0.006 cm$^{-3}$, (B) the standard view of a comet long tail extending 1000's of AU.

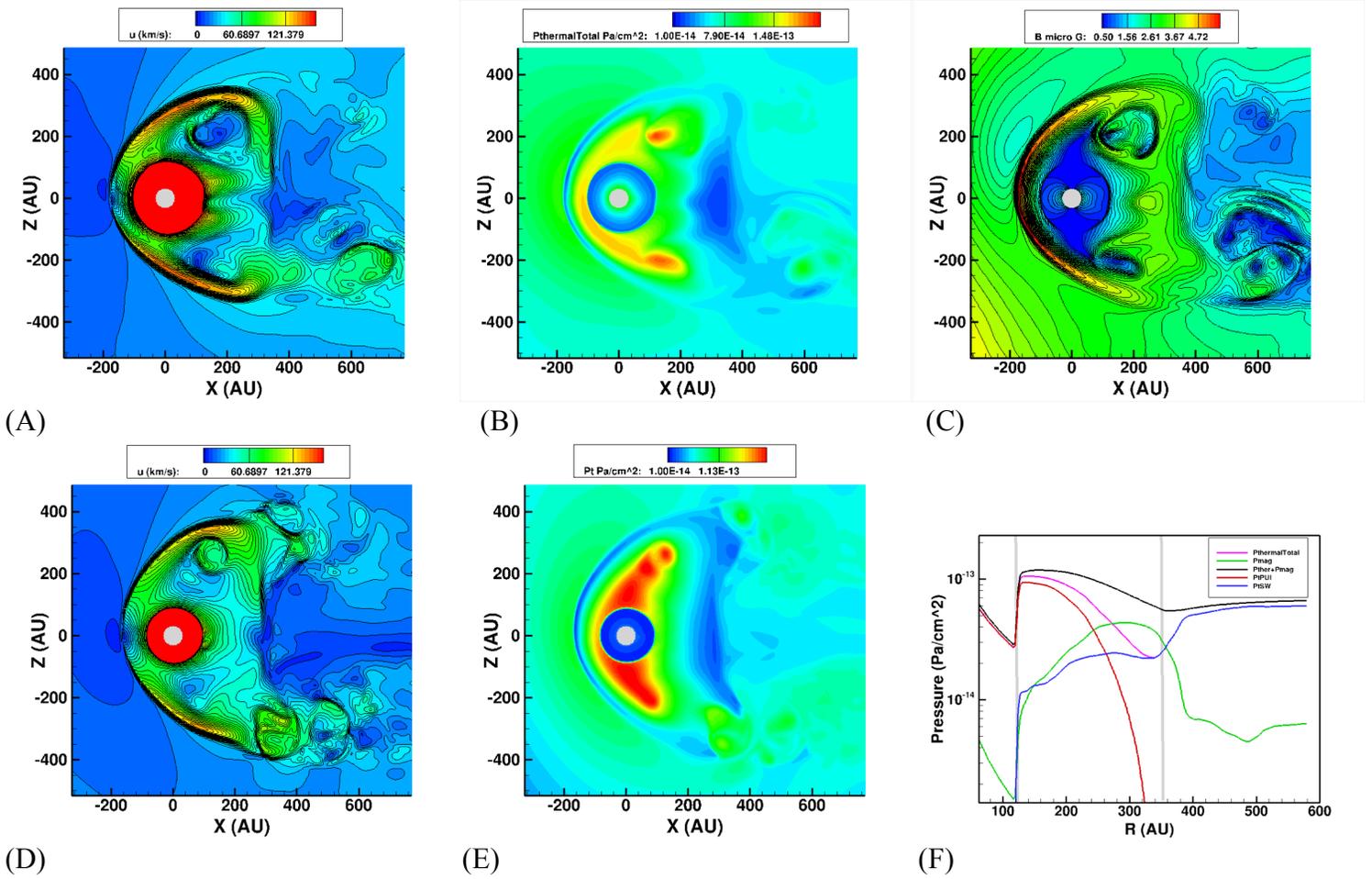

**Figure 2. Meridional Cuts showing the difference when PUIs and thermal ions are treated as separate fluids or not.** Speed (left column), thermal Pressure (middle column) and magnetic field (right column) for the case when PUI and thermal ions are treated as separate fluids (Case A) (A) (B) and (C) panels and for the single ion (D) and (E). In panel (F) show the pressures in the tail along a cut downstream (at z=0). The red line is the PUI pressure; the blue line is the solar wind thermal pressure, the magenta line is the total thermal pressure (PUI + SW); the green line is the magnetic pressure and the black the total pressure (thermal + magnetic). The two gray vertical lines denote the positions of the TS and the heliopause.

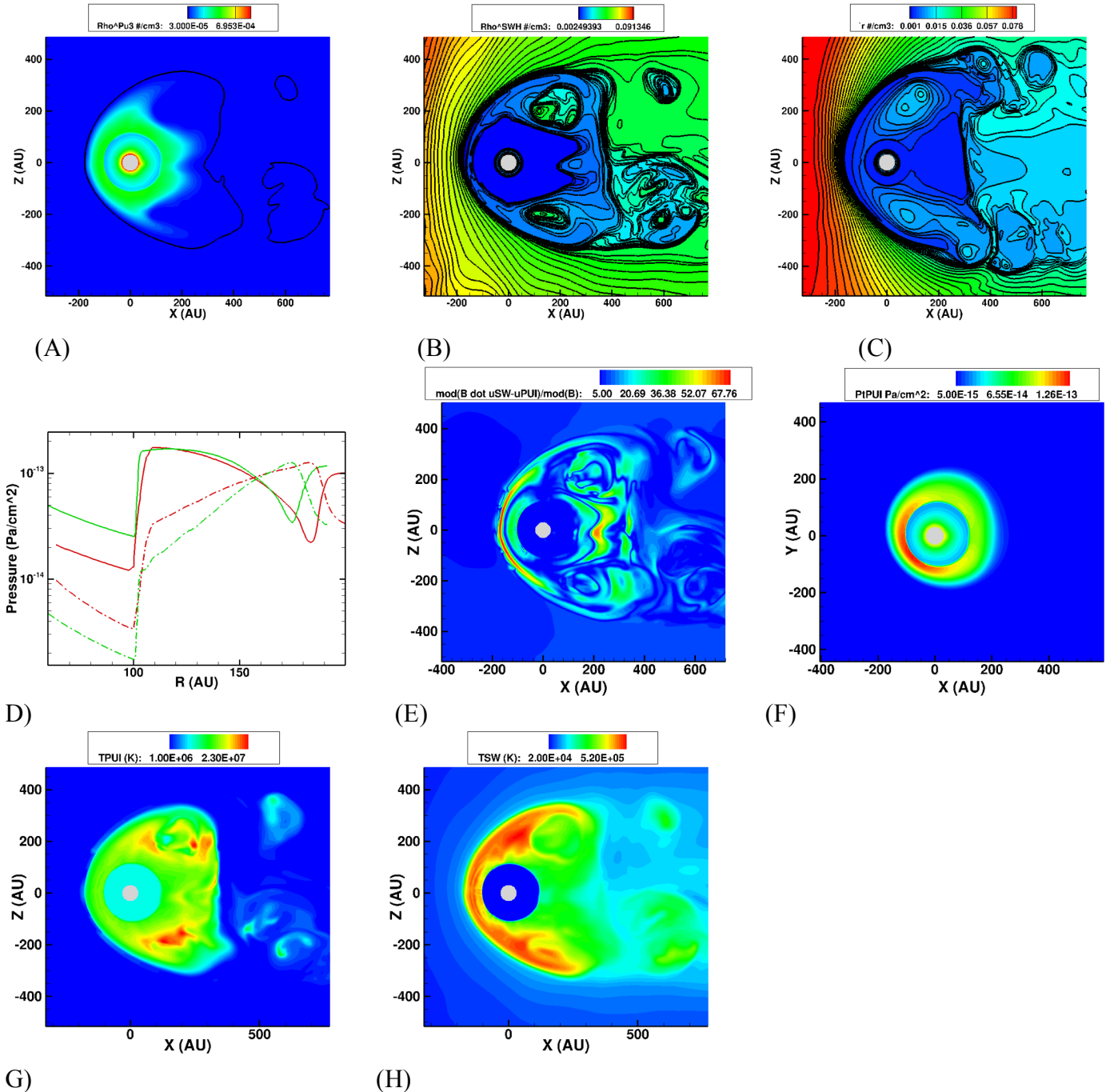

(A) (B) (C)

D) (E) (F)

G) (H)

**Figure 3. Density of PUIs and Solar Wind.** Panel (A) shows the density of PUIs (line contour is temperature at 0.25MK indicating the heliopause.) (B) density of solar wind; (C) density of the single fluid ion (PUI and solar wind combined) model (D) upstream cut showing in green the case with multi-ion model; in red the single ion model; the full lines are the thermal pressure and the dashed lines the magnetic pressure. In the multi-ion case (model A) the thermal pressure is the total thermal pressure of the PUIs and the solar wind. Note that the single ion case was shifted by 21AU; (E) Field aligned velocity difference between PUIs and solar wind = $|B \cdot (u_{SW} - u_{PUI})|/|B|$ in km/s; (F) Thermal pressure of the PUIs in the equatorial plane. (G) Temperature of PUI; (H) Temperature of SW.

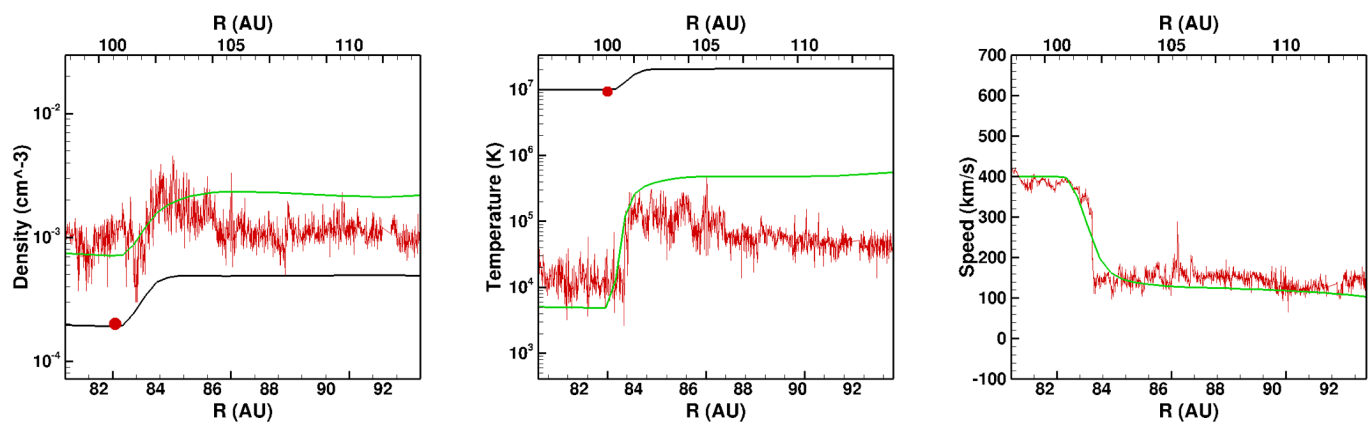

(A)                                        (B)                                        (C)

**Figure 4: Termination Shock Crossing at Voyager 2.** (A) Density; (B) Temperature and (C) Speed. Green line is the thermal solar wind component; black is the PUI component and red are the V2 measurements. The red dot indicates the values predicted upstream the Termination Shock based on the measurements of New Horizon[11]. The bottom axis is the radial distance from the Sun as measured by V2 and the top axis as measured by the model (Model A).

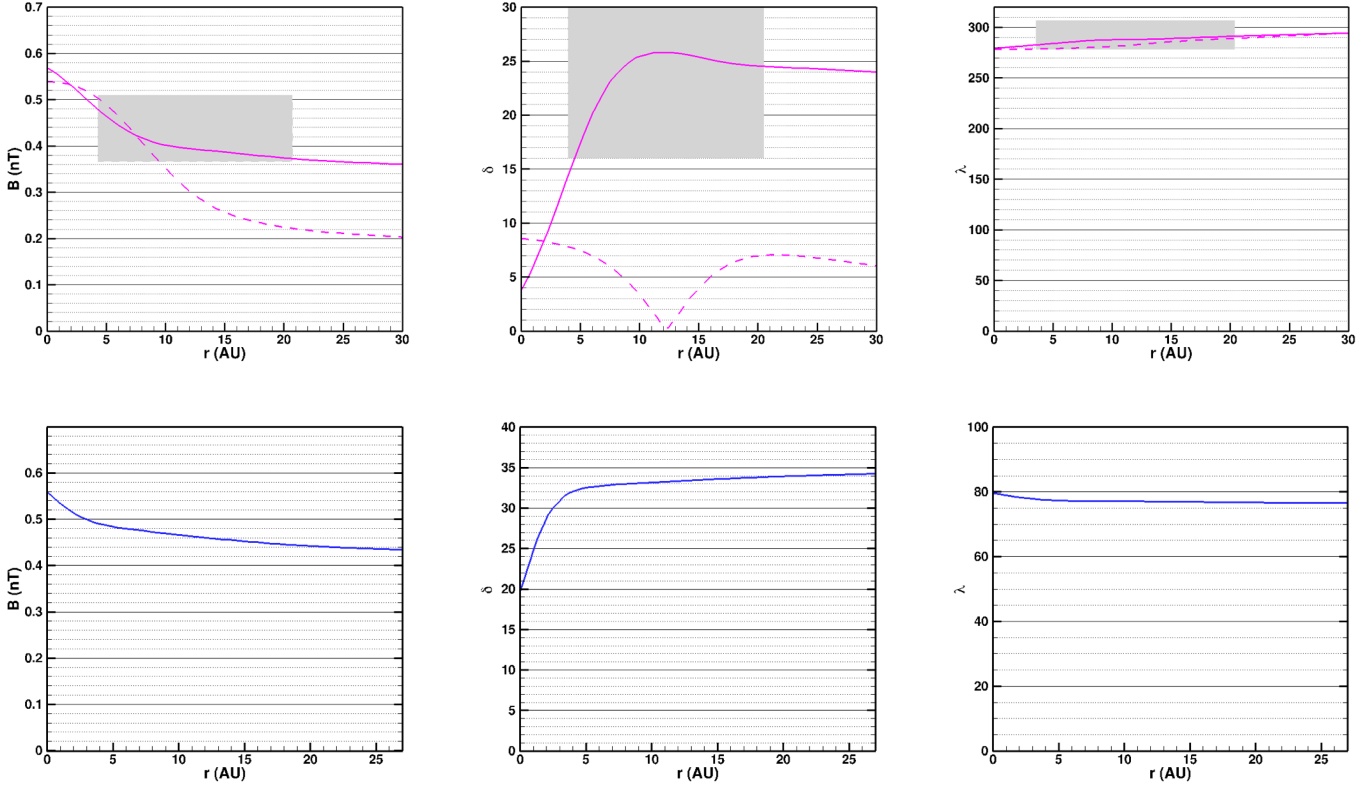

(A)                                      (B)                                    (C)

**Figure 5: Magnetic Field Outside Voyager 1 and 2.** Panel (A) shows the angle $\delta = sin^{-1}(B_N/B)$; panel (B) $\lambda = tan^{-1}(B_T/B_R)$ and (C) the magnitude of the magnetic field, where the *RTN* coordinate system is the local Cartesian system centered at the spacecraft. *R* is radially outward from the Sun, *T* is in the plane of the solar equator and is positive in the direction of solar rotation, and *N* completes a right-handed system for Model A (full line) and B (dashed line). The gray boxes are the observations[14] for Voyager 1 (top) and Voyager 2 (bottom). The variables are plotted vs the distance outside the HP at V1 and V2. The fast rise in angle $\delta$ for the first 10AU after the HP is due to interstellar magnetic field line causally connected (by Alfvén waves) to the solar magnetic field at the eastern flank by reconnection[34].

**Table 1: Distances to Termination Shock (TS), Heliopause (HP) and thickness of Heliosheath (HS)**

| | Model A | | Model B | |
|---|---|---|---|---|
| | **Single Ion** | **Multi Ion** | **Multi ion** | **Observations** |
| TS (V1) | 85AU ± 3AU | 105AU ± 3AU | 102AU ± 3AU | 95AU |
| HP (V1) | 187AU ± 3AU | 190AU ± 3AU | 162AU ± 3AU | 122AU |
| HS (V1) | 102AU | 85AU | 60AU | 28AU |
| | | | | |
| TS (V2) | 80AU ± 3AU | 100AU ± 3AU | 98AU ± 3AU | 85AU |
| HP (V2) | 162AU ± 3AU | 173AU ± 3AU | 157AU ± 3AU | 119AU |
| HS (V2) | 82AU | 73AU | 59AU | 35AU |
| HS (V1-V2) | 20AU | 12AU | 1AU | 7AU |
| | | | | |
| TS (upwind) | 82AU ± 3AU | 85AU ± 3AU | 87AU ± 3AU | - |
| TS (downwind) | 92AU ± 3AU | 95AU ± 3AU | 100AU ± 3AU | - |